\documentclass{article}
\usepackage{spconf,amsmath,graphicx}
\usepackage{calc,amsfonts,amssymb,amsmath,bm,url,color,theorem,graphicx,cite}
\usepackage{psfrag,float}
\usepackage{algorithm}
\usepackage{algorithmic}
\usepackage{subcaption}

\definecolor{orange}{RGB}{255,107,0}

\newtheorem{Assumption}{Assumption}
\newtheorem{Fact}{Fact}

\newtheorem{Corollary}{Corollary}

\theorembodyfont{\rmfamily}

\newtheorem{Remark}{Remark}

\def\va{{\bm a}}

\def\vu{{\bm u}}

\def\vs{{\bm s}}
\def\vx{{\bm x}}

\def\vh{{\bm h}}
\def\vq{{\bm q}}

\def\mH{{\bm H}}

\def\mD{{\bm D}}

\def\jj{{\mathfrak{j}}}

\def\va{{\bm{a}}}

\def\vd{{\bm{d}}}

\def\vg{{\bm{g}}}
\def\vh{{\bm{h}}}

\def\vq{{\bm{q}}}

\def\vs{{\bm{s}}}

\def\vu{{\bm{u}}}

\def\vx{{\bm{x}}}

\def\mD{{\bm{D}}}

\def\mH{{\bm{H}}}

\def\jj{{\mathfrak j}}

\newcommand{\C}{\mathbb C}

\title{Transmitting Data Through Reconfigurable Intelligent Surface: A Spatial Sigma-Delta Modulation Approach
}
\name{Wai-Yiu Keung${}^{\dagger \mathsection}$, Hei Victor Cheng${}^{\ddagger}$, Wing-Kin Ma${}^{\dagger}$ \vspace{-5pt}}
\address{\small
${}^{\dagger}$Department of Electronic Engineering, The Chinese University of Hong Kong, Hong Kong SAR of China \\ \small
${}^{\mathsection}$Department of Computer Science and Engineering, The Chinese University of Hong Kong, Hong Kong SAR of China\\\small
${}^{\ddagger}$Electrical and Computer Engineering Department, Aarhus University, Denmark\vspace{-10pt}
}

\begin{document}
\ninept
\allowdisplaybreaks
\setlength{\belowdisplayskip}{3pt} \setlength{\belowdisplayshortskip}{3pt}
\setlength{\abovedisplayskip}{3pt} \setlength{\abovedisplayshortskip}{3pt}
\maketitle
\begin{abstract}
Transmitting data using the phases on reconfigurable intelligent surfaces (RIS) is a promising solution for future energy-efficient communication systems. Recent work showed that a virtual phased massive multiuser multiple-input-multiple-out (MIMO) transmitter can be formed using only one active antenna and a large passive RIS. In this paper, we are interested in using such a system to perform MIMO downlink precoding. In this context, we may not be able to apply conventional MIMO precoding schemes, such as the simple zero-forcing (ZF) scheme, and we typically need to design the phase signals by solving optimization problems with constant modulus constraints or with discrete phase constraints, which pose challenges with high computational complexities. In this work, we propose an alternative approach based on Sigma-Delta ($\Sigma\Delta$) modulation, which is classically famous for its noise-shaping ability. Specifically, first-order $\Sigma\Delta$ modulation is applied in the spatial domain to handle phase quantization in generating constant envelope signals. Under some mild assumptions, the proposed phased $\Sigma\Delta$ modulator allows us to use the ZF scheme to synthesize the RIS reflection phases with negligible complexity. The proposed approach is empirically shown to achieve comparable bit error rate performance to the unquantized ZF scheme.

\end{abstract}
\vspace{-5pt}
\section{Introduction}\vspace{-5pt}
\label{sec:intro}
Reconfigurable intelligent surface (RIS), also known as intelligent reflective surface (IRS), has garnered significant attention for its capability to enhance the spectral and energy efficiencies of existing communication systems \cite{reflecting_metasurface2017,basar2019wireless,liang2019large,huang2019reconfigurable}. RIS consists of an array comprising numerous adjustable reflective elements, where each element can introduce phase shifts to incoming electromagnetic waves. Through precise control of the phase shifts, the RIS can manipulate the electromagnetic wave to achieve various outcomes, such as constructive signal reflection towards intended destinations \cite{wu2019intelligent,jiang2020}, or nullification of signals at specific locations to counter interference \cite{tao2022}.
What sets RIS apart is its passive nature, enabling it to operate with minimal energy consumption when inducing the phase shifts. This characteristic makes it a cost-effective alternative to the traditional relay systems \cite{emil2020}. Given these advantages,  RIS applications have proliferated in the design of diverse wireless communication systems in the literature. However, most of the existing work employs RIS solely as a passive beamformer. In this scenario, the phase shifts at the RIS are solely determined by channel realizations. It is worth noting that passive beamforming does not fully exploit the potential capabilities of the RIS, as shown in \cite{dof_isit2021}.

The central focus of this paper lies in the exploration of a novel use case for the RIS, where the phase shifts at the RIS are not solely determined by the channel but also incorporate information data. When data is made accessible at the RIS, it can be used to modulate the information by adjusting the phase shifts, thereby further enhancing the overall channel capacity. The modulation of information using RIS entails conveying data by modifying the transmission environment. Building upon this concept, it becomes possible to establish a virtual massive multiple-input-multiple-output (MIMO) system using just a single active transmit antenna at the base station (BS) in conjunction with an RIS with a large number of reflective elements. This configuration empowers the simultaneous provision of downlink data to multiple users, owing to the multiplexing gain described in \cite{cheng2021}. This architecture greatly reduces the requirement for radio-frequency components at the base station and simplifies the digital signal processing at the BS. Consequently, this design represents an energy-efficient solution for future communication systems.
	
The concept of modulating information through the transmission environment has been previously explored, referred to as ``media-based communication'' \cite{media_modulation}. In the setup of using RIS to encode information, earlier work \cite{karasik2021} focuses on deriving channel capacity under fixed finite modulation for joint information transmission. Information encoding through RIS has also been addressed in \cite{yan2020} and \cite{reflecting_modulation}, where the objective was to design algorithms to jointly decode the information contained in the transmitted signals and the reflective coefficients. In \cite{cheng2021}, the multiplexing gain associated with joint transmission from the transmitter and the RIS is characterized. The results demonstrated the potential for substantial improvement in the multiplexing gain. A solution involving symbol-level precoding is proposed in \cite{cheng2022slp}, showcasing the attainability of this multiplexing gain. However, this approach requires solving optimization problems with constant modulus constraints, the high computational complexity renders the idea infeasible in practice. 

Thus far, despite the attractiveness of the idea of using RIS to transmit information, there is no practical design available. This paper focuses on filling this gap by proposing a low-complexity optimization-free solution based on Sigma-Delta ($\Sigma\Delta$) modulation. Spatial $\Sigma\Delta$ modulation has recently emerged as a new paradigm for quantized MIMO, in both uplink channel estimation and downlink precoding \cite{rao2019-sigdel-icassp, shao2019-onebit, shao2020-2ndorder}. As a classical technique used in analog-to-digital/digital-to-analog converter (ADC/DAC), $\Sigma\Delta$ modulation features quantization noise-shaping. When applied to one-bit MIMO downlink, spatial $\Sigma\Delta$ modulation allows a simple zero-forcing precode-then-quantize strategy by containing the quantization error experienced by the target users \cite{shao2019-onebit}. Motivated by this, our interest lies in studying how spatial $\Sigma\Delta$ modulation can be applied to design the reflective coefficients for the implementation of data-transmitting RIS. We will show that $\Sigma\Delta$ modulation provides a low-complexity solution for information modulation through the environment. \vspace{-5pt}
% ===========================================================================
\section{Problem Settings}
\begin{figure}[t]\vspace{-10pt}
	\centering
\includegraphics[width=0.99\linewidth]{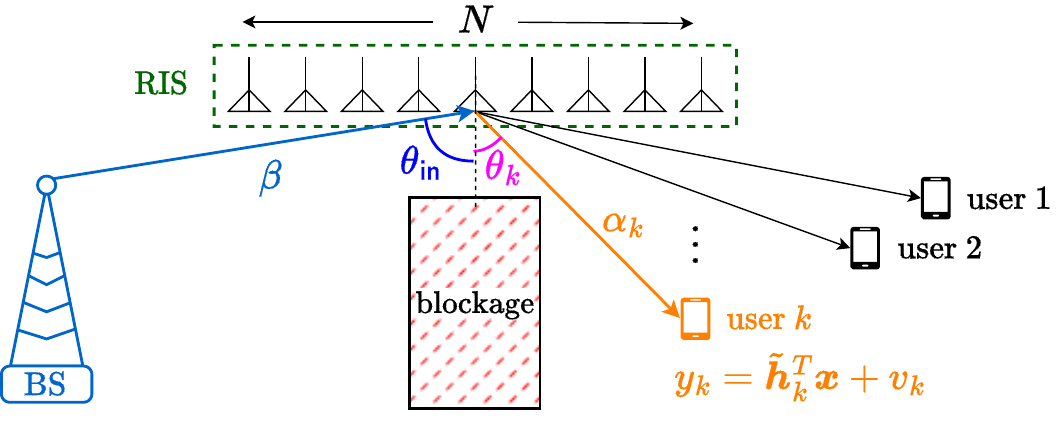}\vspace{-10pt}
	\caption{Scenario.}\vspace{-10pt}
	\label{fig:scenario}
\end{figure}
 Consider a RIS-assisted downlink scenario using a BS with only one active antenna, where the BS only provides the required power and the phases on RIS are used to modulate information. The problem settings of concern are illustrated in Figure~\ref{fig:scenario}. Assuming the RIS is a uniform linear array (ULA), the received signal for user $k$ can be written as
 \begin{equation}\label{eqn:system-model}
     y_k = \Tilde{\vh}_k ^\top  \vx + v_k = (\vg \odot \vh_k)^\top  \vx + v_k
 \end{equation}
where 
\[
    \vg = \beta(1, e^{-\jj 2\pi \frac{d}{\lambda} \sin(\theta_{\sf in})}, \dots, e^{-\jj 2\pi \frac{d}{\lambda} \sin(\theta_{\sf in})(N-1)})
\]
is the arrival steering vector from the BS to the RIS; and 
\begin{subequations}\label{eqn:downlink-steering}
 \begin{align} 
    \vh_k &= \alpha_k \va(\theta_k)\\
    \va(\theta_k) &= (1, e^{-\jj 2\pi \frac{d}{\lambda} \sin(\theta_{k})}, \dots, e^{-\jj 2\pi \frac{d}{k} \sin(\theta_{k})(N-1)})
\end{align}   
\end{subequations}
is the downlink steering vector of the RIS to the $k$-th user. Observe the resultant end-to-end channel from the BS to the user:
\begin{align*}
    \Tilde{\vh}_k &= \underbrace{(\beta \alpha_k)}_{\mbox{$\Tilde{\alpha}_k$}} \cdot \underbrace{(1, e^{-\jj \omega}, \dots , e^{-\jj \omega (N- 1)})}_{\mbox{$\boldsymbol{\Tilde{a}}$} }, \\ \omega &= 2\pi ({d}/{\lambda}) [\sin(\theta_{\sf in})+ \sin(\theta_k)] .
\end{align*}
Remark that this is also referred to as a cascaded channel. Here, $\theta_{\sf in}$ is the angle of arrival from the BS to the RIS, and $\theta_k$ is the angle of departure from the RIS to the $k$-th user. The background noise $v_k$ is assumed to follow complex Gaussian distribution with zero mean and $\sigma_v^2$ variance. We assume the direct link between the BS and the user is unavailable because, e.g. there is a blockage between the source and sink. The constant envelope signal vector
\[
\vx = (e^{\jj \psi_1}, \dots, e^{\jj \psi_N})
\]
are the RIS's reflection coefficients to be designed. In practice, the reflection angles $\{\psi_n\}_{n = 1}^N$ should be quantized to a finite number of phases. 
They are reconfigured with respect to the channel state information and the carried information available at a centralized processing unit that controls both the BS and the RIS. In the current literature, many of the existing works focus on using optimization over $\vx$, or equivalently,  $\boldsymbol{\psi} = (\psi_1, \dots, \psi_N)$, to achieve a satisfying performance metric, e.g. the sum rate or the minimum rate of the BS and the $k$-th receiver. 
We assume the centralized processing unit has full knowledge of the CSI by, such as channel estimation through the pilot stage, in the remainder of this paper.  \vspace{-5pt}

%===========================================================================
\section{Spatial $\Sigma\Delta$ Modulation for RIS}
% ===========================================================================
\subsection{First-order $\Sigma\Delta$ Modulation}
\begin{figure}[t]\vspace{-10pt}
	\centering
\includegraphics[width=0.9\linewidth]{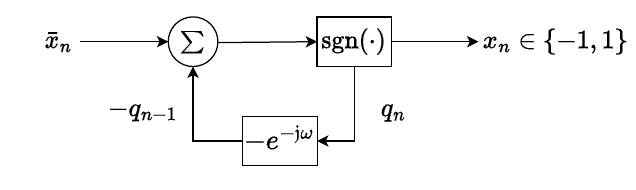}\vspace{-10pt}
	\caption{System diagram of the first-order $\Sigma\Delta$ modulator.} \vspace{-10pt}
	\label{fig:sigdel-temp}
\end{figure}

 In this section, we briefly review the concept of $\Sigma\Delta$ modulation \cite{aziz-1996-overview-spmag}. Figure~\ref{fig:sigdel-temp} shows the system diagram of the first-order $\Sigma\Delta$ modulator. Widely used in temporal ADC/DAC designs, the system reads in a real-valued sequence $\bar{x}_n$ and generates a binary sequence by 
\begin{equation} \label{eqn:sigdel-temp}
 x_n = {\rm sgn}(\Bar{x}_n - q_{n-1}) = \Bar{x}_n - q_{n-1} + q_n,
\end{equation}
where $q_n$ denotes the quantization distortion introduced at the signum function ${\rm sgn}(\cdot)$, which is used to model an one-bit converter. The term $q_{n-1}$ denotes the delayed quantization noise, which is fed back to the quantizer's input. By taking Fourier transform on both sides of \eqref{eqn:sigdel-temp}, the modulator can be described by the system equation
\begin{equation}\label{eqn:DTFT-sigdel}
    X(\omega) = \bar{X}(\omega) + (1 - e^{-\jj \omega}) Q(\omega),
\end{equation}
where $U(\omega)$ denotes the discrete time Fourier transform of the sequence $u_n$. Notice the noise-shaping response $G(\omega) = 1-e^{-\jj \omega}$ is a high-pass filter. This implies the quantization noise $q_n$ is shaped toward the high frequency band. By assuming the input signal $\bar{x}_n$ is low-pass, it can be separated from the quantization noise by applying a low-pass filter at the one-bit output signal. This technique is called \textit{noise-shaping} in the $\Sigma\Delta$ literature. 
We should mention that the $\Sigma\Delta$ modulator may suffer from unbounded quantization noise due to the feedback path of $q_{n- 1}$. A sufficient condition to avoid this is to limit the input signal amplitude to $|\Bar{x}_n| \leq 1$, which leads to a bound on $|q_n| \leq 1$ \cite{gray-1990-quantnoisespectra}. This is called the no-overload condition. Effectively, this guarantees that the quantization noise $q_n$ in the system \eqref{eqn:sigdel-temp} is stable.  
\vspace{-
0pt}

In \cite{shao2019-onebit}, $\Sigma\Delta$ modulation has been applied in the spatial domain to handle binary MIMO downlink precoding. The key assumption is that the base station is a uniform linear array (ULA), which means the downlink channel follows \eqref{eqn:downlink-steering}. We summarize some important aspects of the space-time duality when applying a spatial $\Sigma\Delta$ modulator to MIMO downlink:
\begin{itemize}
    \item the temporal feedback in \eqref{eqn:sigdel-temp} is treated as passing the quantization noise incurred at each antenna to the adjacent antenna;
    \item the noise-shaping effects in the spatial frequency domain, i.e. the angle relative to the ULA's broadside; and
    \item the low-pass nature of the temporal input signal leads to the requirement on the user's angle to be small, e.g. within $\pm 20^\circ$.  
\end{itemize}
Furthermore, the band-pass $\Sigma\Delta$ modulation \cite{Schreier1989Bandpass, aziz-1996-overview-spmag} adjusts the noise-shaping response $G(\omega)$ to null a designated frequency. When applied to spatial $\Sigma\Delta$ modulation, the effect is that the quantization noise will shaped toward a sector region other than near the broadside, e.g., $G(\omega)$ can be a notch filter that has zero response at $30^\circ$. This is proposed as an angle-steered $\Sigma\Delta$ modulator \cite{shao2019-onebit}, as to be described in the next session.

% ===========================================================================
\subsection{Angle-Steered $\Sigma\Delta$ Modulator with Phase Quantizer}
Consider an angle-steered spatial $\Sigma\Delta$ modulator that generates discrete constant envelope output alphabet with $L$ points sampled on the unit circle $\mathcal{X} = \{e^{\jj \frac{2\pi \ell}{L}}, \ell = 1, \dots, L\}$. The system diagram is shown in Figure~\ref{fig:sigdel}. The $\Sigma\Delta$ phase-quantization process can be characterized as
\begin{equation}
    x_n = \mathcal{Q}_L(\Bar{x}_n - e^{\jj \phi} q_{n-1}), \label{eqn:sigdel}
\end{equation}
the phase-quantizer $\mathcal{Q}_L(a): \C \mapsto \mathcal{X}$ returns $a$ to the nearest phase on the unit circle, i.e. $x_n \in \mathcal{X}$, and $q_n$ is the quantization error introduced by $\mathcal{Q}_L$. The end-to-end relation of the modulator is 
\begin{equation} \label{eqn:sigdel-end2end}
    x_n = \Bar{x}_n + q_n - {e^{\jj \phi}} q_{n-1}.
\end{equation}   
The effect of the phase-shifting term $e^{\jj \phi}$ in the feedback path of $q_{n-1}$ lies in tilting the noise-shaping effect from the low-pass region to a region centering at $\phi$, as can be seen from the system response 
\begin{equation*}
    X(\omega) = \bar{X}(\omega) + (1 - e^{-\jj (\omega + \phi)}) Q(\omega),
\end{equation*}
following \eqref{eqn:DTFT-sigdel}. The noise-shaping response is now a notch filter that is null at $\phi$. To ensure the error sequence magnitude $|q_n|$ is bounded, we make an assumption on the input sequence $x_n$ using: 
\begin{figure}[t]\vspace{-0pt}
	\centering
\includegraphics[width=1.0\linewidth]{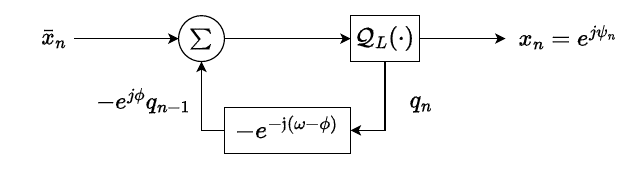}\vspace{-10pt}
	\caption{The angle-steered $\Sigma\Delta$ modulator.}\vspace{-10pt}
	\label{fig:sigdel}
\end{figure}

\begin{figure}[t!] 
\begin{minipage}[c]{\linewidth}
%  \centering\vspace{6pt}
  %\centerline{\includegraphics[width=\textwidth]{wyFig/diffNoUsrs.eps}}
    \begin{subfigure}[t]{0.5\textwidth}
       \centerline{\includegraphics[width=\textwidth]{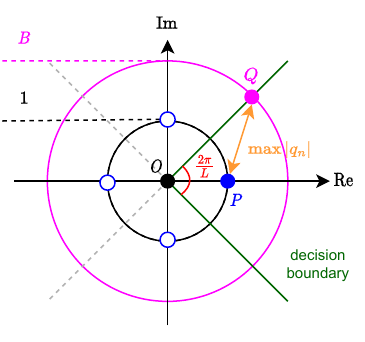}}\vspace{-0pt}
       \caption{The Complex Plane.} \label{fig:sketchproof-a}
\end{subfigure}\hfill
\begin{subfigure}[t]{0.4\textwidth}
       \centerline{\includegraphics[width=\textwidth]{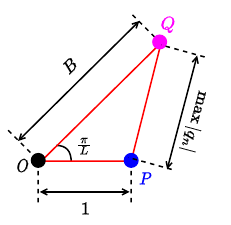}}\vspace{-0pt}
       \caption{Triangle $\triangle OPQ$}\label{fig:sketchproof-b}
\end{subfigure}
\end{minipage}
\caption{Sketch Proof of Fact 1.}  \label{fig:sketchproof} \vspace{-10pt}
\end{figure}

\begin{Fact} Consider the $\Sigma\Delta$ modulator with a discrete phase quantizer $\mathcal{Q}_L(\cdot)$ in \eqref{eqn:sigdel}, or in Figure~\ref{fig:sigdel}, using a discrete phase quantizer with $L \geq 4$, where $L$ is an integer. Suppose the input sequence $\bar{x}_n \in \C$ satisfies  $|\Bar{x}_n| \leq A_L^\star$, where 
\begin{equation} \label{eqn:no-overload}
    A_L^\star = \frac{\sin(2\pi / L)}{\sin(\pi/L)} - 1 \leq 1,
\end{equation}
then, the quantization noise sequence $q_n$ satisfies $|q_n| \leq 1$.    
\end{Fact}\vspace{-5pt}
\noindent{\it Proof.}  Let $b_n = \Bar{x}_n - e^{\jj \phi}q_{n-1}$ be the quantizer input, and denote $B = \max_{n} |b_n|$ such that $|q_n| \leq 1$. Observe the triangle formed by the origin $O = (0 , 0)$, the phase-quantized point $P = (1, 0)$ and the point $Q$ lying on the circumference of the circle originating at $O$ and with radius $B$, and intersects the decision boundary of $\mathcal{Q}_L$. See a visualization of the complex plane in Figure~\ref{fig:sketchproof-a}. If we fix $\max |q_n| = \overline{QO} = 1$, we see $\triangle OPQ$ in Figure~\ref{fig:sketchproof-b} is an isosceles triangle.\footnote{We fix $\max |q_n| = 1$ because it is the worst case maximum quantization noise when $B \geq 0$, which corresponds to the quantizer's input of $b_n = 0$.} Thus, we can construct the sine law $B/\sin(\pi - 2\pi/L) = 1/\sin(\pi/L)$, 
which can be rewritten as 
\begin{align} \label{eqn:B_max}
    B  &= \frac{\sin(\pi -2\pi/L)}{\sin(\pi/L)}
    %&= \frac{\sin(\pi) \cos(2\pi/L) - \cos(\pi) \sin(2\pi/L)}{\sin(\pi/L)} \\
    = \frac{\sin(2\pi/L)}{\sin(\pi/L)},
\end{align}
wherein we have used the identity $\sin(\pi - \alpha) = \sin(\alpha)$. The last inequality follows from $\sin(2\pi/L)/\sin(\pi/L)=2\cos(\pi/L)\leq 2$ using the double angle formula for sine function. The proof is completed by observing the no-overload condition $|\bar{x}_n| \leq B - 1$.  \hfill $\blacksquare$\vspace{-5pt}
% ============================================================================
\begin{Corollary} Consider the $\Sigma\Delta$ modulator in \eqref{eqn:sigdel}. The maximum modulator amplitude is $|\bar{x}_n| \leq A = 1$ when $L \to \infty$, i.e. when the quantizer is continuous-phased. 
\end{Corollary}
{\it Proof.} It suffices to check, in \eqref{eqn:B_max}, $B = 2$ when $L \to \infty$. The ratio 
\[
     \lim_{L \to \infty}\frac{\sin(2\pi/L)}{\sin(\pi/L)} =  \lim_{L \to \infty}\frac{2\pi/L}{\pi/L} = 2
\]
wherein we used $\lim_{\alpha \to 0}\sin(\alpha)/\alpha = 1$, which is true when $L$ is large. %together with the face that $\sin(2\pi/L)/\sin(\pi/L)=2\cos(\pi/L)$ is a monotonic increasing function in $L$ completes the proof. 
\hfill$\blacksquare$\vspace{-5pt}

\begin{Remark}[Overloading] We should remark that Fact 1 is a sufficient condition to guarantee $|q_n| \leq 1$. However, it is also true that overloading does not necessarily lead to unbounded quantization noise. In fact, it has been reported that a moderate level of overloading could lead to performance gains in empirical observations, e.g. in \cite{shao2019-onebit, shao2020-2ndorder}. Precisely, it is quite common in the $\Sigma\Delta$ literature to purposefully violate the input sequence amplitude constraint \eqref{eqn:no-overload} by feeding in a signal that has amplitudes larger than $A_L^\star$. We resort to empirical studies in order to illustrate the effect overloading may potentially bring.  
\end{Remark}

We now discuss the principle of $\Sigma\Delta$ precoding in the context of interest. Given a free-space precoding vector $\boldsymbol{\Bar{x}}\in \C^N$, which can be generated from traditional precoding scheme such as the zero-forcing precoder, we design the reflective coefficients $\vx \in \mathcal{X}^N$ by spatial $\Sigma\Delta$ modulation. The received signal model yields 
\[
y_k = \Tilde{\vh}_k^\top (\boldsymbol{\bar{x}} + \vq - \vq^-) + v_k = \Tilde{\alpha}_k \boldsymbol{\Tilde{a}}^\top \boldsymbol{\Bar{x}} + w_k
\]
wherein we denote \(\vq = (q_1, q_2, \dots, q_N)\) as the quantization error vector, and  \(\vq^- = (0, q_1, \dots, q_{N-1})\) as the delayed quantization error vector. The effective noise component $w_k$ can be evaluated as 
\begin{align*}
    w_k &= \Tilde{\alpha}_k \boldsymbol{\Tilde{a}}^\top(\vq-\vq^{-}) + v_k,
\end{align*}
and the inner product is 
\begin{align*}
    \boldsymbol{\Tilde{a}}^\top(\vq-\vq^{-}) &= \sum_{n = 1}^{N- 1} (q_n - q_{n-1})e^{-\jj \left(\phi - 2\pi \frac{d}{\lambda} [\sin(\theta_{\sf in}) + \sin(\theta_k)]\right)n}\\
    &= (1- z^{-1})\sum_{n = 0}^{N- 2} q_{n + 1} z^{-n} + z^{-(N-1)}q_N ,
\end{align*}
where we let $z = e^{\jj \left(\phi -  2\pi \frac{d}{\lambda} [\sin(\theta_{\sf in}) + \sin(\theta_k)] \right)}$. Using the same derivations in \cite[eqn. 11]{shao2019-onebit}, we assume a large $N$ and obtain
\[
\sigma_w^2 \approx {2|\Tilde{\alpha}|}{\sigma_q^2} \left|\sin\left(\frac{\phi - \frac{2\pi d}{\lambda} [\sin(\theta_{\sf in}) + \sin(\theta_k)]}{2}\right)\right|^2 + \sigma_v^2
\]
where $\sigma_q^2 = E(|q_n|^2)$ is the variance of $q_n$, and $\sigma_v^2$ is the background noise power. This means we can choose $\phi^\star = \frac{2\pi d}{\lambda} [\sin(\theta_{\sf in}) + \sin(\theta^\star)]$ such that $\theta_k \in [\theta^\star-\delta, \theta^\star + \delta]$ in order to minimize $\sigma_w^2$. The physical meaning of the above derivation is that, if the user angles $\theta_k$'s lie within a small angular sector, then the quantization error term $q_n$ will have little effect on the effective noise power $\sigma_w^2$. The following assumption is used in the above derivations:
\begin{Assumption} The magnitude of the quantization error sequence $|q_n|$ is uniformly i.i.d. in  $[-1, 1]$, and the phase $\angle q_n$ is uniformly distributed on $[-\pi, \pi]$. This allows us to write $\sigma_q^2 = 1/3$. 
\end{Assumption}

We consider $\Sigma\Delta$-ZF precoding\cite[Subsection 5.1]{shao2019-onebit} for the design of $\boldsymbol{\bar{x}}$ in the following. Specifically, the precoding vector reads
\[
\boldsymbol{\Bar{x}}_t = C \boldsymbol{\Tilde{A}}^\dagger \mD \vs_t, \qquad C = \frac{ A_L^\star }{\max_{n ,t} |[\boldsymbol{\Tilde{A}}^\dagger \mD \vs_t]_n|}
\]
whereas $\mD = {\sf diag}(\vd),$ $\vd = (d_1, \dots, d_K)$, and $d_k = {\sigma_w \Tilde{\alpha}_k^*}/{|\Tilde{\alpha}_k|^2}$. Under this setting, the received signals read as 
\[
    y_k = C \sigma_{w,k} \cdot s_k + w_k,
\]
and the symbol detection task at the receiver side can be performed by $\hat{s}_k = {\sf dec}(y_k/(C\sigma_w))$. \vspace{-5pt}

\begin{Remark}[Subtractive Dithers] It is worth-noting that Assumption 1 may not hold in general as the quantization error is, ultimately, dependent on the input signal sequence $\bar{x}_n$. A widely adopted technique in combating this reality is to introduce some subtractive dithers. Specifically, given an artificially generated random dither signal $\vu = (u_1, \dots, u_N)$, where $|u_n| \leq A_L^\star$, we modify the input signal by, e.g.
\begin{equation}\label{eqn:dithers}
    \boldsymbol{\Bar{x}}_t = (0.8C) \boldsymbol{\Tilde{A}}^\dagger \mD \vs_t + (0.2)\vu.
\end{equation}
At the receiver side, we can subtract the dither back by re-modifying the decision function as 
\[
\hat{s}_k = {\sf dec}\left(\frac{y_k - 0.2\Tilde{\vh}_k^\top \vu}{0.8C\sigma_w}\right)
\]
to counteract the effect of dithering on the detection performance. We can do so because the user devices can obtain the CSI $\Tilde{\vh}_k$ and the scaling term $C\sigma_w$ from pilot stage, and $\vu$ can be regenerated at the user's side by using the same random seed. 
\end{Remark}\vspace{-10pt}

\section{Numerical Results}
This section evaluates the bit error rate (BER) performance of the proposed $\Sigma\Delta$-ZF, the direct quantized ZF and the unquantized ZF. Precisely, the unquantized ZF makes an ideal assumption on the RIS such that it can absorb different levels of energy and reflect 
\begin{equation} \label{eqn:zero-forcing}
    \vx^{({\sf ZF})}_t = \Tilde{\mH}^\dagger \vs_t\Big/\max_{n, t}|[\Tilde{\mH}^\dagger \vs_t]_n|. 
\end{equation}
This implies $\vx^{({\sf ZF})}_t \in \{\vx \in \C^N \,|\, \|\vx\|_\infty \leq 1\}$, which is not a constant envelop signal vector. 
The quantized ZF means the RIS reflects only the angle of the ZF precoder, i.e. $\vx = \mathcal{Q}_L(\vx^{({\sf ZF})}_t)$, which satisfies the constant envelop nature for the RIS reflective coefficients. 

The simulation settings are as follows. The number of antenna element at the RIS is $N = 512$, and the number of target remote users is $K = 8$. The inter-antenna spacing of the RIS is $d = \lambda/8$. The angle of arrival from the BS to the RIS is $\theta_{\sf in} = -60^\circ$. The complex channel gain $\beta$ has a phase uniformly drawn from $[-\pi, \pi]$, and the amplitude is generated by $|\beta| = r_0/r_1$, where $r_0 = 30$ and $r_1$ is uniformly drawn from $[20, 100]$. The angle of departure from the RIS to users are randomly picked from $\theta_k \in [20^\circ, 60^\circ]$, and they are separated by at least $1^\circ$. The complex channel gains from the RIS to users $\alpha_k$'s are generated in the same fashion of that for $\beta$. The results are obtained by $1000$ Monte-Carlo trials of block-fading channels, with the block length $T = 500$. We assume a symbol constellation of $16$-ary QAM in our simulations. 

\begin{figure}[t]
	\centering
\includegraphics[width=0.9\linewidth]{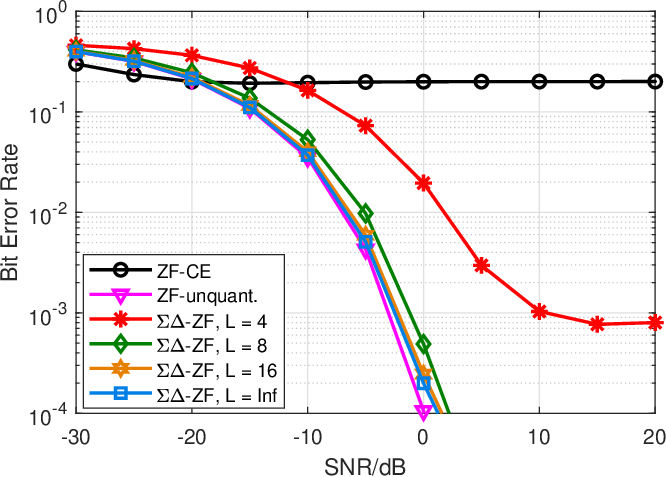}
	\caption{BER performance of the $\Sigma\Delta$-ZF scheme under different number of discrete phases $L$.}\vspace{-8pt}
	\label{fig:sigdel-diff-L}
\end{figure}
Figure~\ref{fig:sigdel-diff-L} shows the BER performance for the proposed $\Sigma\Delta$-ZF method with different number of discrete phases $L$. The benchmark scheme ``CE-ZF" means the ZF scheme \eqref{eqn:zero-forcing} is constant-envelop quantized with continuous phase, i.e. $\vx = \mathcal{Q}_\infty(\vx^{({\sf ZF})}_t)$. The result indicates that the proposed $\Sigma\Delta$-ZF scheme may achieve the performance of unquantized ZF with a higher number of discrete phases $L$. Also, it is noticed that the performance gain between $L = 4$ and $L = 8$ is substantial, but the gain of increasing from $L = 8$ to $L = 16$ is smaller. The performance of $L = 16$ is comparable to the continuous phase counterpart.

\begin{figure}[t]
	\centering
\includegraphics[width=0.9\linewidth]{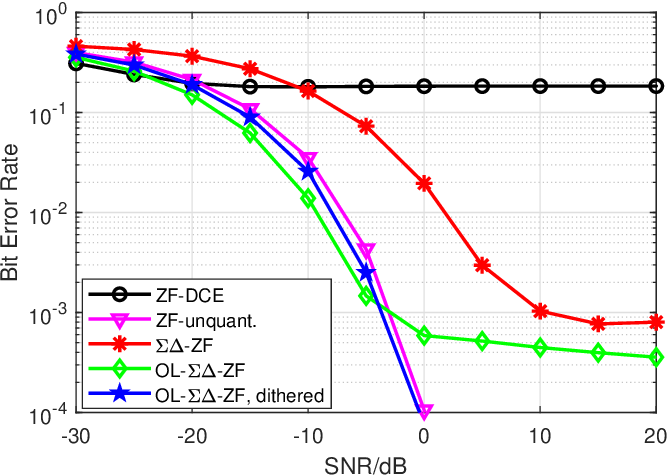}
	\caption{BER performance of the $\Sigma\Delta$-ZF scheme under overloading and subtractive dithers; $L = 4$.}\vspace{-18pt}
	\label{fig:sigdel-ol-dith}
\end{figure}

In the previous experiment, we found that the $\Sigma\Delta$-ZF scheme with $L = 4$ discrete phases admits a huge performance gap with the unquantized ZF counterpart, as it suffers from heavy error flooring effects. We try to narrow that gap by using the heuristics of overloading and subtractive dithers. Figure~\ref{fig:sigdel-ol-dith} presents the simulation results. The legend ``ZF-DCE" means the $4$-phases quantized signal of the unquantized ZF \eqref{eqn:zero-forcing}, i.e. $\vx = \mathcal{Q}_4(\vx^{({\sf ZF})})$. Here, we first overload the $\Sigma\Delta$ modulator by intentionally increasing the input signal amplitude to $A = \sqrt{2}$, which violates \eqref{eqn:no-overload} in Fact 1. The performance is captured by the legend of ``OL-$\Sigma\Delta$-ZF". We see an improvement in the low SNR region, but the effect error flooring remains severe. Next, we apply subtractive dithers \eqref{eqn:dithers} to the overloaded modulator. The result is captured by the legend ``OL-$\Sigma\Delta$-ZF, dithered". It is empirically found that the proposed scheme with $L = 4$ achieves a comparable BER performance with the unquantized ZF, under the assistance of the tricks in the aforementioned remarks. 
\vspace{-4pt}
\section{Conclusions}
This paper investigates the practical design of using phases on RIS for transmitting information. It provides an optimization-free solution based on applying $\Sigma\Delta$ modulation to RIS reflective coefficient designs, which is known to be hard due to the constant modulus constraints or discrete phase constraints. The proposed $\Sigma\Delta$ approach is computationally friendly and showcases competitive BER performance in numerical simulations. More practical settings, such as when the RIS is a rectangular planar array, and when there exist direct links between the BS and the users, are left for future studies.

\newpage 
\bibliographystyle{IEEEbib}
\bibliography{ref}

\end{document}